\documentclass[12pt]{article}

\usepackage[body={17.5cm, 22.5cm},right=2cm]{geometry}

\usepackage{float}

\usepackage{booktabs}
\usepackage{caption}

\usepackage{color}
\usepackage{graphicx}
\usepackage{epsf}
\usepackage{graphicx,epsfig}
\usepackage{amsmath}

\usepackage{multirow}

\usepackage{amsmath, amsthm, amssymb}

\usepackage{epsfig}
\usepackage{cite}
\usepackage{color,colordvi}

\newcounter{hran}

\makeatletter
\renewcommand\section{\@startsection {section}{1}{\z@}%
                               {-3.5ex \@plus -1ex \@minus -.2ex}%
                               {2.3ex \@plus.2ex}%
                               {\normalfont\large\bfseries}}
\makeatother

\begin{document}\thispagestyle{empty}

\vspace{0.5cm}

\def\thefootnote{\arabic{footnote}}
\setcounter{footnote}{0}

\def\s{\sigma}
\def\nn{\nonumber}
\def\p{\partial}
\def\ls{\left[}
\def\rs{\right]}
\def\lc{\left\{}
\def\rc{\right\}}
\def\oT{\overline{T}}
\def\t{\tau}
\def\s{\sigma}

\newcommand{\be}{\begin{eqnarray}}
\newcommand{\ee}{\end{eqnarray}}
\newcommand{\bi}{\begin{itemize}}
\newcommand{\ei}{\end{itemize}}
\renewcommand{\th}{\theta}
\newcommand{\bth}{\overline{\theta}}
\newcommand{\avg}[1]{\ensuremath{\left\langle #1 \right\ranlge}}

\hspace*{12cm}
CERN-PH-TH/2014-046

\vspace*{2cm}

\begin{center}

{\Large \bf 
The Imaginary Starobinsky Model
 }
\\[1.5cm]
{\normalsize   \large Sergio Ferrara$^{1,2,3}$,  Alex Kehagias$^{4,5}$ and Antonio Riotto$^{5}$
}
\\[1.1cm]

\vspace{.1cm}
{\small {  \it $^{1}$ Physics Department, Theory Unit, CERN,
CH 1211, Geneva 23, Switzerland}}
\\

\vspace{.1cm}

\vspace{.1cm}

{${}^2$\sl\small INFN - Laboratori Nazionali di Frascati \\
Via Enrico Fermi 40, I-00044 Frascati, Italy}
\\

\vspace{.2cm}
{${}^3$\sl\small Department of Physics and Astronomy,
University of California \\ Los Angeles, CA 90095-1547, USA}
\\

\vspace{.2cm}

{\small { \it $^{4}$ Physics Division, National Technical University of Athens, 15780 Zografou Campus, 
Athens, Greece}}\\

\vspace{.2cm}
{\small {  \it $^{5}$ Department of Theoretical Physics
24 quai E. Ansermet, CH-1211 Geneva 4, Switzerland}}\\




\end{center}

\vspace{1.2cm}

\begin{center}
{\small  \noindent \textbf{Abstract}} \\[0.5cm]

\end{center}
\noindent 
{\small
 The recent detection by the BICEP2 collaboration  of a high level of tensor modes  seems to exclude the Starobinsky model of inflation. In this paper we show that  this conclusion can be avoided: one can   embed the Starobinsky model in supergravity and identify the inflaton field  with the imaginary (instead of the real) part of the chiral scalaron multiplet in its formulation. Once coupled to matter, the Starobinsky model may then become the chaotic quadratic model with shift symmetry during inflation and is in good agreement with the
current data.
\vskip 2cm

\def\thefootnote{\arabic{footnote}}
\setcounter{footnote}{0}





\baselineskip= 19pt
\newpage 
\section{Introduction}\pagenumbering{arabic}

The recent Planck results \cite{ade} have indicated that the cosmological perturbations in the Cosmic Microwave
Background (CMB) radiation are nearly gaussian and of the adiabatic type. If one insists in assuming
that these scalar perturbations are to be ascribed to single-field model of inflation \cite{lr}, the data put severe
constraints restriction on the inflationary  parameters. In particular, the Planck results have strengthened the upper
limits on the tensor-to-scalar ratio, $r< 0.12 $ at 95\% C.L., disfavouring many inflationary models.
In particular, the
simplest quadratic chaotic model has been excluded at about 95\% C.L.

Among the inflationary models discussed by the Planck collaboration is the Starobinsky $(R + R^2)$ theory, 
 first presented in Refs. \cite{star} (see also  Ref. \cite{Whitt:1984pd}).
The Starobinsky model   is a model that leads to a quasi-de Sitter phase and it is described  by the Lagrangian 

\be
 S_{\rm S}=\frac{M_{\rm p}^2}{2}\int {\rm d}^4 x\sqrt{-g} \ \left(R+\frac{1}{6M^2} R^2\right), \label{star}
 \ee
 where $M_{\rm p}$ is the reduced Planck mass.
 This theory  does not describe only the GR degrees of freedom, {\it i.e.} the helicity-2 massless graviton, but in addition it propagates a scalar degree of freedom  usually called "scalaron". The later is hidden in the action (\ref{star}) and can be revealed  in the so-called linear representation, where  one writes 
the action  in the equivalent form \cite{Whitt:1984pd}
\be
\label{R2}
 S_{\rm S}=\int {\rm d}^4 x\sqrt{-g} \ \left(\frac{M_{\rm p}^2}{2}R + \frac{M_p}{M}  R\psi-
 3 \psi^2\right).
 \ee
After integrating out the field $\psi$, one gets back the original theory (\ref{star}). However, the action (\ref{R2}) is written in a Jordan frame and it can be expressed in Einstein frame after the conformal transformation
\be
g_{\mu\nu}\to e^{-\sqrt{2/3} \phi/M_{\rm p}}g_{\mu\nu}=\left(1+\frac{2\psi}{M M_{\rm p}}\right)^{-1} g_{\mu\nu} 
\ee
is performed. Then, 
we get the equivalent scalar field version of the Starobinsky model 
\be
 S_{\rm S}=\int {\rm d}^4 x\sqrt{-g}\left[\frac{M_{\rm p}^2}{2}R-\frac{1}{2}\partial_\mu\phi\partial^\mu \phi-\frac{3}{4} M_{\rm p}^2 M^2\left(1-e^{-\sqrt{\frac{2}{3}}\phi/M_{\rm p}}\right)^2\right]. \label{R3}
\ee
There is a plateau in the scalar potential for large values of $\phi$ where slow-roll inflation can be realised with a quasi-de Sitter phase driven by a vacuum energy 
\be 
V_{\rm S}=\frac{3}{4}M_{\rm p}^2 M^2.  \label{VS}
\ee
The normalization of the CMB anisotropies  fixes  $M\approx 10^{-5}M_{\rm p}$. 
In addition, the  scalar tilt $n_S$ and  tensor-to-scalar ratio $r$ turns out to be
\be
\label{pred}
n_S-1
\approx
-\frac{2}{N},\, \,\,r\approx \frac{12}{N^2}. 
\ee
Note that $r$ has an addition $1/N$ suppression  with respect 
to $n_S$. Although this model looks quite ad hoc at the theoretical level, it is perfect agreement with the Planck data, 
 basically due to an  additional $1/N$ suppression ($N$ being the number of e-folds till the end of inflation)
of $r$ with respect to the prediction for the scalar spectral index $n_S$. 

For this reason, there has been a 
a renewed interest on the Starobinsky   model, with  particular emphasis  on its supergravity  extensions\cite{KL,sugrastaro,ENO,FKLP,FKR,FKD,KT}, along the lines originated in Refs. \cite{fv,Cecotti:1987sa,CFPS}.

This positive attitude versus the Starobinsky model has dramatically changed with the recent release of the measurement of the tensor
modes from large angle CMB B-mode polarization by
BICEP2 \cite{bicep}, implying  a tensor-to-scalar ratio 

\be 
r=0.2^{+0.07}_{-0.05}.
\ee 
Putting aside the tension with the Planck data,  this result (if confirmed)   puts inflation  on a ground which is  firmer than ever. On the other side, it is in contradiction with the predictions (\ref{pred}) of the Starobinsky model.

The goal of this paper is to show that this is not necessarily true: the contradiction with the tensor modes data disappear if one embeds
the Starobinsky model in supergravity and identifies the inflaton field with the  imaginary part of the chiral multiplet in the dual formulation of the model (instead of the  the real part of it, as done in all the literature so far). 
We dub this version of the Starobinsky theory the ``Imaginary Starobinsky model" and show that
it basically resembles
 the quadratic chaotic model during inflation \cite{chaotic} (for recent reviews, see \cite{Kallosh:2014ona}) once the coupling to matter is considered (a necessary condition to allow reheating in the model).
 It is nice  that just 
 embedding  the Starobinky model into supergravity can make  it in agreement with the data. 
 
Recently in \cite{KLr} it was shown that the simplest proposal in the standard supersymmetric Starobinsky model to identify the axion $b$, partner of the scalaron, with inflaton does not work and a drastic modification to the theory must be made if the $b$ field is
responsible for the inflation. We show that a plausible modification can be made which naturally leads to $b$ inflation.

The paper is organized as follows. In section 2 we recall the basics of the embedding of the Starobinsky model in supergravity, as done in the literature so far. In section 3 we describe our proposal
to identify the inflaton with the partner of the ``scalaron" rather than the scalaron itself. Section 4 contains
the main points about the imaginary Starobinsky model. Section 5 contains our conclusions.

\section{The Supergravity embedding of Starobinsky model }

The bosonic Starobinsky model  can be embedded in ${\cal N}=1$ minimal supergravity. In fact, since it is a higher curvature theory, it can be described both in old-minimal \cite{Cecotti:1987sa} as well as in new-minimal \cite{CFPS} ${\cal N}=1$ supergravity. 

\subsection{Inflaton potential embedded in new-minimal supergravity}

The $(R+R^2)$ gravity dual, in the (new-minimal) off-shell formulation of supergravity corresponds to a massive vector multiplet $(1,2(1/2),0)$ with a self-coupling function $J(C)$ where $C$ is the scalar partner of the massive vector. In this scenario, the D-term potential is
$g^2 (J'(C))^2$ and the metric for the $C$-field is just $-J''(C)$ ($J''(C)<0)$. In the Starobinsky model dual to the $(R+R^2)$ supergravity we have (in $M_{\rm p}=1$ units)
\begin{eqnarray}
 J(C)=\frac{3}{2}\big{[}C+\ln(-C)\big{]},~~~ C=-e^{\sqrt{\frac{2}{3}}\phi}. \label{j}
 \end{eqnarray} 
It is only when $J(C)$ is given by (\ref{j}) that the model reproduce pure $(R+R^2)$ supergravity. In all other cases, the $(R+R^2)$ theory is coupled to an extra massive vector multiplet \cite{FKLP}. This is similar to the findings of \cite{CK} in 
old-minimal formulation when we depart from a superpotential $W=ST$ to a new superpotential $SF(T)$, where $S,T$ are the chiral superfields of the old minimal formulation. 
 
This theory is also equivalent to (new-minimal) standard supergravity coupled to a linear multiplet and a gauge field which makes the linear multiplet massive \cite{CFG}. 
The mass terms for the vector is 
 $m_B^2=-J''(C) B_\mu^2$
 so if we choose a free vector coupled to gravity, $J''(C)={\rm const.}$ and the D-term just generates the mass term for the vector
 \begin{eqnarray}
 V_D=\frac{g^2}{2}C^2.
 \end{eqnarray}
Note in this model the F-I term is irrelevant because of the invariance of the kinetic term under $C\to C+\xi$. 

The $\alpha$-models \cite{FKLP} are obtained by the K\"ahler potential (of the complex field $z$ where ${\rm Im}\, z$ has been eaten by the vector)   $K=-3 \alpha\ln C$ so that the canonically normalized field becomes $C=-\exp \sqrt{\frac{2}{3\alpha}}\phi$. The curvature of $SU(2,R)/U(1)$ coset is now \cite{FKLP} 
\be
R_{z\bar{z}}=-\frac{2}{3\alpha}.
\ee
For each $\alpha$ there are actually three models one can construct but only one, corresponding  to the gauging of the parabolic isometry, gives the potential 
\be
V=\frac{g^2}{2} \left(1-e^{-\sqrt{\frac{2}{3\alpha}}\phi}\right)^2.
\ee
Other gaugings produce potentials with $\sinh\sqrt{\frac{2}{3\alpha}}\phi$ or $\cosh\sqrt{\frac{2}{3\alpha}}\phi$ \cite{FFS}. The $\alpha\to \infty $ model reproduce the Stuckelberg 
model \cite{KLR}, while the $\alpha\to 0$ model reproduce  the Freedman model \cite{F0} (where the $U(1)$ symmetry is restored and the potential just becomes a cosmological constant).

\subsection{Inflaton potential embedded in old-minimal supergravity}

The  Starobinsky Lagrangian is usually embedded \cite{Cecotti:1987sa} in the   ``old minimal'' two--derivative formulation of supergravity with  the gravitational supermultiplet  coupled to a pair of additional chiral multiplets, the inflaton $T$ and the goldstino $S$ multiplets. In the minimal universal embedding, supergravity is actually coupled   to  a chiral multiplet containing the inflaton and a goldstino multiplet $X$ replacing $S$ \cite{ADFS}. The latter is a constraint superfield \cite{goldstino} obeying 
\be
X^2=0.
\ee 
Such a constraint superfield has been used before for inflation \cite{luis}.
This constraint in fact allows to solve the scalar of the chiral multiplet 
in terms of the goldstino bilinear $GG$ and $X$ is explicitly  expressed as 
\begin{equation}
X \ = \ \frac{GG}{2 \, F_X} \ + \ \sqrt{2}\, \theta\, G \ +\  \theta^2 F_X. \label{va1}
\end{equation}
The supergravity Lagrangian is then written in the conformal compensator formalism \cite{SUGRA} as
\begin{equation}
{\cal L} \ = \ - \ \bigg[ \big(T \,+ \,{\overline T} \,- \, |X|^2\big) S_0 \, {\overline S}_0 \bigg]_D \ +\
\bigg[ \big(M X T \,+ \,f X \,+\, W_0\big) S_0^3 \, + \, {\rm h.c} \bigg]_F. \label{d1}
 \end{equation}
By using the identity 

\be
\bigg[ (T + {\overline T}) S_0 \, {\overline S}_0\bigg]_D \ = \ \bigg[T \, {\cal R} \, S_0^2\bigg]_F \ + \ {\rm h.c.},
\ee
where $ {\cal R}$ is the chiral supergravity multiplet, (\ref{d1})  can be expressed as
\begin{equation}
{\cal L} \ =  \ \bigg[ |X|^2 \,S_0 \, {\overline S}_0 \bigg]_D \ +\
\left[ \left(T\big(-\,\frac{\cal R}{S_0}\,+ \, M \, X\big)  \,+ \,f \,X \,+ \,W_0 \right) S_0^3 \ + \ {\rm h.c} \right]_F.
 \label{d2}
 \end{equation}
 Let us note that $T$ appears in \eqref{d2} as a Lagrange multiplier, and its equation of motion is simply
\begin{equation}
X \ =  \ \frac{1}{M}\frac{\cal R}{S_0}. \label{d3}
\end{equation}
Due to the $X^2 =0$ constraint, the chiral supergravity mutiplet ${\cal R}$ satisfies also
\begin{equation}
{\cal R}^2 = 0 . \label{d4}
\end{equation}
This  constraint can be implemented by a chiral Lagrange multiplier $\sigma$ and the dual action to (\ref{d1}) turns out to be 

\begin{equation}
e^{-1}{\cal L} \ = \ - \ \left[ S_0 \, {\overline S}_0 \, - \, \frac{{\cal R} {\cal {\overline R}}}{M^2} \right]_D
\ + \ \left[ W_0\, + \, \xi \, \frac{{\cal R}}{S_0} \ S_0^3 \, + \, \sigma \, {\cal R}^2 \, S_0 \right]_F. \label{d5}
\end{equation}
Then, by recalling that the components of the chiral superfield ${\cal R}$ are
\cite{SUGRA,fv,fkvp,FKP}
\begin{equation}
{\cal R} = \left( {\overline u} \equiv S \,+\, i P\, , \ \gamma^{mn} {\cal D}_m \psi_n \, ,\ - \ \frac{1}{2}\ R
\,-\, \frac{1}{3} \ A_m^2 \,+\, i \,{\cal D}^m A_m \,- \,\frac{1}{3} \ u \, {\overline u}  \right), \label{s1}
\end{equation}
where $u$ and $A_m$ are the  auxiliary fields of ``old-minimal" $N=1$ supergravity and $\psi_n$ is the gravitino field, we find that the bosonic Lagrangian of (\ref{d5}) is \cite{ADFS}
\begin{equation}
{\cal L} \ = \ \frac{1}{2}\ \left(R\, + \, \frac{2}{3}\ A_m^2\right)  \ + \ \frac{3}{4 M^2} \ \left(R\,+ \, \frac{2}{3}\ A_m^2\right)^2 \  +  \ \frac{3}{M^2} \ ({\cal D}_m A^m)^2, \label{d15}
\end{equation}
which clearly describes an $(R+R^2)$ supergravity coupled to a pseudoscalar mode coming from ${\cal D}_m  A^m$ . 

We now proceed with the dual action (\ref{d1}) and   a simple inspection of  it shows that the 
 K\"ahler potential and superpotential are given  by 
\be
\label{K1}
K = -3\, \ln\Big{(}T + \overline{T} - X \overline{X}\Big{)} 
\ee
and 
\be
\label{W1}
W=  3 \sqrt{\lambda} \, X T +f X +W_0, 
\ee%
respectively. 
Although  the goldstino  superfield $X$ is not dynamical as it does not contains any elementary scalar field, it contributes to the scalar potential since (at $X=0$)
\be
F_{\overline{X}} \ = \ { e^{\frac{K}{2}}} \left( K_{X\,\overline{X}} \right)^{\,-\,1} \ {\overline{W}}_{\overline{X}}.
\ee
 The  scalar potential is then given by 
\begin{equation}
V \ = \ \frac{|M T \,+ \, f|^2}{3 \, ( T \, + \, {\overline T})^2}, \label{vas2}
\end{equation}
where $M=3\sqrt{\lambda}$ \cite{FKR}
and the 
bosonic Lagrangian turns out to be
\begin{equation}
e^{-1}{\cal L} \ = \ \frac{R}{2} \ - \ \frac{3}{(T \,+ \, {\overline T})^2} \ |\partial \,T|^2\  - \ \frac{|M \,T\, + \, f|^2}{3 ( T \,+ \,{\overline T})^2}. \label{vas3}
\end{equation}
%
Note that the positivity of $V$ in (\ref{vas2}) is due to the no-scale structure of the $T$-inflaton K\"ahler potential \cite{CF}.

It is standard to identify the inflaton with the real part of the complex scalar $T$. Indeed, parametrizing  the scalaron $Re(T)$  as 
 \begin{eqnarray}
 {\rm Re}\, T= e^{\sqrt{\frac{2}{3}} \phi}, 
 \end{eqnarray}
 and integrating out the ${\rm Im}\, T$, 
 we find that the effective bosonic theory,  after appropriate shift of $\phi$, turns out to be
 
\be
\label{OM7}
e^{-1}{\cal L}_1 = \frac{1}{2} R  -  \frac{1}{2} \partial_{\mu} \phi  \partial^{\mu} \phi 
-\ \frac{3}{4 } \,\lambda  \left(1-e^{-\sqrt{\frac{2}{3}} \phi}\right)^2.
\ee
This is  the standard Starobinsky model in the dual theory. However, as we have already mentioned, it cannot account for the level of the gravitational waves indicated by  BICEP2. As a result, one is tempting to rule out also the supersymmetric $(R+R^2)$ theory. However, unlike the non-supersymmetric case, the supersymmetric $(R+R^2)$ theory
has a solution encoded in itself. This is the subject of the next section.

\section{The Imaginary Supersymmetric Starobinsky Model}
As we have seen above, identifying the inflaton with the real part of the $T$-field does not provide a large enough amount of tensor modes. However, the field  $T$ has  also an imaginary part, which we have fixed before to its vacuum value ${\rm Im}\,T=0$. In the general case of a complex $T$,
 
\begin{eqnarray}
 V=3\lambda \frac{|T+c|^2}{(T+\bar{T})^2} \, ,  \label{VF2}
 \end{eqnarray} 
where $c=f/3\sqrt{\lambda}$, such that the theory is now described by the Lagrangian 

{\begin{eqnarray}
  e^{-1} {\cal L}=\frac{1}{2} R -3\frac{|\partial T|^2}{(T+\bar{T})^2}-
  3\lambda \frac{|T+c|^2}{(T+\bar{T})^2}.
  \end{eqnarray}
 It is dual to the $(R+R^2)$ theory described in (\ref{d15}).
 After parametrizing  
the complex scalar $T$ by two real scalar $\phi$ and $b$, as

\be
\label{T}
T=e^{\sqrt{\frac{2}{3}} \phi} + i\sqrt{\frac{2}{3}}\,  b, 
\ee
we find that  the effective bosonic theory turns  out to be 

\be
\label{OM71}
e^{-1}{\cal L} = \frac{1}{2} R  -  \frac{1}{2} \partial_{\mu} \phi  \partial^{\mu} \phi -\frac{1}{2} e^{-2\sqrt{\frac{2}{3}} \phi}\partial_\mu b \partial^\mu b-\frac{1}{2}\lambda e^{-2\sqrt{\frac{2}{3}} \phi}\,  b^2
-\ \frac{3}{4} \lambda \,  
\left(1-\frac{1}{2}e^{-\sqrt{\frac{2}{3}} \phi}\right)^2.
\ee
Since the
field $\phi$ is present both in the kinetic and in the mass term of the field $b$, 
we can consider instead of $b$, the new field  $\chi=e^{-\sqrt{\frac{2}{3}} \phi}\, b$. We consider now the initial configuration where the $\phi$ field is close to the minimum of its potential. Its 
energy density is (still in Planck units)  $\ll \lambda\sim 1$ and it is much smaller than the one associated to the
$\chi$ field, which is $\sim \chi^2\gg 1$. Therefore 
the energy density of the $\phi$ field is completely negligible at the beginning. However, as it has been shown in Ref. \cite{KLr,jap},  even for an initial  large value of the $\chi$ the latter  is immediately driven  to oscillate around $\chi=0$ and inflation ends almost instantaneously. The reason for this
behavior is  the kinetic mixing of the fields $\chi$ and $\phi$. Their kinetic terms are of the form

\be
{\cal L}_{\rm kin}=\frac{1}{2}\dot\phi^2\left(1+\frac{2}{3}\chi^2\right)+\frac{1}{2}\dot\chi^2+\sqrt{\frac{2}{3}}\chi\dot\chi\dot\phi
\ee
and  the equations of motion of the fields are

\begin{eqnarray}
\frac{{\rm d}}{{\rm d} t}\left(\dot\chi+\sqrt{\frac{2}{3}}\chi\dot\phi\right) +3H\left(\dot\chi+\sqrt{\frac{2}{3}}\chi\dot\phi\right)+\left(\lambda-\frac{2}{3}\dot\phi^2\right)\chi-\sqrt{\frac{2}{3}}\dot\chi\dot\phi&=&0,\nonumber\\
\frac{{\rm d}}{{\rm d} t}\left[\dot\phi\left(1+\frac{2}{3}\chi^2\right)+\sqrt{\frac{2}{3}}\chi\dot\chi\right]+3H\left[\dot\phi\left(1+\frac{2}{3}\chi^2\right)+\sqrt{\frac{2}{3}}\chi\dot\chi\right]+\frac{3}{8}\sqrt{\frac{2}{3}} \lambda \,  e^{-\sqrt{\frac{2}{3}\phi}}
\left(1-\frac{1}{2}e^{-\sqrt{\frac{2}{3}} \phi}\right)
&=&0.\nonumber\\
&&
\ee
For initial  values $\chi\gg 1$, the field $\phi$ is pushed towards the plateau of its potential and its equation is solved by  $\dot\phi\simeq -\sqrt{3/2}\dot\chi/\chi$, thus canceling the friction term of the field $\chi$ and making the latter rapidly rolling
to the minimum of its potential. From this discussion we can infer that one needs to strongly stabilize the field 
$\phi$ with a large curvature around the minimum of its potential. This can be achieved by 
considering  the couplings of the Starobinsky model to matter. These couplings are a necessary ingredient for the Starobinsky  model in order to let the universe reheat after the end of inflation. 
Therefore, one should include  couplings of the multiplet $T$ to matter multiplets $\Phi_i$. As explained in Ref.  \cite{EKN}, we will consider that these couplings are induced by modifying the K\"ahler potential  as
\be
K=-3 \ln\left(T+\oT-X\overline{X}+ (T+\oT)^n F(\Phi_i)+{\rm h.c.}\right)+ K_m(\Phi_i,\overline{\Phi}_i).
\ee
Assuming that  all matter scalars are stabilized at 
$\langle\Phi_i\rangle$ with 
$\langle  D_iW\rangle=0$ and $F(\langle \Phi_i\rangle)=m$,  the dynamics of the $T$ and $X$ multiplets are then effectively described by the  K\"ahler potential 
\be
K=-3 \ln\left(T+\oT-X\overline{X}+ m (T+\oT)^n\right). \label{kah}
\ee
Different modification of the K\"ahler potential have been considered in \cite{KLr} and shown to stabilize the $\phi$ field during inflation triggered by the b field.
%
We also assume that  the superpotential $W(T,X,\Phi_i)$ is such that it takes the standard form
\begin{eqnarray}
W(T,X)=W(T,X,\langle\Phi_i\rangle)=  3 \sqrt{\lambda} \, X T +f X .
\end{eqnarray}
The resulting potential then turns out to be
\begin{eqnarray}
 V=3\lambda \frac{|T-f|^2}{\left[T+\bar{T}+m (T+\overline{T})^n\right]^2}  \label{VF2m}
 \end{eqnarray}
and the K\"ahler metric is 
\begin{eqnarray}
K_{T\oT}=3\frac{1+mn(T+\oT)^{n-2}\Big{[}(3-n)(T+\oT)+m(T+\oT)^n\Big{]}}{\left[T+\oT+m (T+\overline{T})^n\right]^2}.
\end{eqnarray}
In terms of the real and imaginary parts of $T$, the potential reads 
\begin{eqnarray}
\label{p}
V(\phi,b)=\frac{3}{4}\lambda \frac{\Big{(}1-f\,e^{-\gamma\phi}\Big{)}^2}{\Big{(}1+2^{n-1}m\,
e^{(n-1)\gamma\phi}\Big{)}^2}+\frac{3}{4}\gamma^2\lambda \frac{e^{-2\gamma\phi}}{\Big{(}1+2^{n-1}m
e^{(n-1)\gamma\phi}\Big{)}^2}b^2  \label{pot}
\end{eqnarray}
where $\gamma=\sqrt{2/3}$,  and the bosonic Lagrangian turns out to be
\begin{eqnarray}
{\cal L}=\frac{1}{2} R-\frac{1+mn (2e^{\gamma\phi})^{n-2}\big{[}
2(3-n)e^{\gamma\phi}+m (2e^{\gamma\phi})^n\big{]}}{2\big{(}1+m (2e^{\gamma\phi})^{n-1}\big{)}^2}\Big{(}\partial_\mu\phi\partial^\mu\phi+e^{-2\gamma\phi}\partial_\mu b\partial^\mu b\Big{)}-V(\phi,b). \label{ll}
\end{eqnarray}
Note that for $m=0$, the 
scalars parametrize the K\"ahler space  $SU(1,1)/U(1)$ and (\ref{ll})    reduces to  (\ref{OM71}). 
For a generic value of $m$, the scalar manifold is deformed such that only a $U(1)$ isometry is preserved. In this case we find that as $m$ tends towards $m=-(2f)^{1-n}$,
the minimum of the potential in the field $\phi$ gets steeper and steeper when $n$ goes to unity. This behaviour is similar with that of the model considered in \cite{KLr}.
\begin{figure}[h]
\centering
\includegraphics[width=80mm]{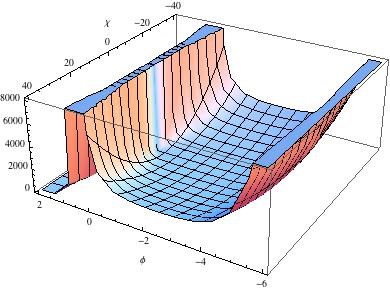}
\caption{The potential (\ref{p}) in terms of the canonically normalized fields and  $n=2$.}
\label{overflow}
\end{figure}

To simplify the discussion, let us take the particular value 
\begin{eqnarray}
\label{point}
 m=-n^{-1}(2f)^{1-n}, 
 \end{eqnarray} 
 with $n\neq 1$, for which 
there is a minimum $\phi=\phi_0=\ln f^{1/\gamma}$ independently of the   value of $b$. The potential is drawn in Fig. 1. The value $n=1$ is excluded as it the K\"ahler potential (\ref{kah}) does not depend on $T$. 
At the minimum

\begin{eqnarray}
\left.\frac{{\rm d}^2 V}{{\rm d}\phi^2}\right|_{\phi_0}=\frac{n^2 \lambda (f^2+\gamma^2n \, b^2)}{f^2(n-1)^2}
\end{eqnarray}
and the field  $\phi$ is anchored there by a large curvature, so that   the potential for the imaginary part of the $T$-field turns out to be
\begin{eqnarray}
V_{\rm eff}(b)=V(\phi_0,b)=\frac{3\gamma^2 n^2\lambda}{4f^2(n-1)^2}\,b^2.
\end{eqnarray}
Since
\begin{eqnarray}
K_{T\oT}\Big{|}_{\phi_0}=\frac{3n}{4f^2} ,
\end{eqnarray}
 the theory  at $\phi=\phi_0$ is described by (with $n>0$ and $n \neq 1$) 
\begin{eqnarray}
{\cal{L}}=\frac{1}{2} R-\frac{3n\gamma^2}{4f^2}\partial_\mu b\partial^\mu b-
\frac{3\gamma^2 n^2\lambda}{4f^2(n-1)^2}\, b^2. \label{l1}
\end{eqnarray}
Upon redefining $\chi=b\sqrt{n}/f$ we write 
the Lagrangian (\ref{l1}) with a canonically normalized kinetic term as 
\begin{eqnarray}
{\cal{L}}=\frac{1}{2} R-\frac{1}{2}\partial_\mu \chi\partial^\mu \chi-
\frac{1}{2}m_\chi^2\chi^2,
\end{eqnarray}
where (since $\lambda=M^2/9)$
\begin{eqnarray}
m_\chi^2=\frac{nM^2}{9(n-1)^2}.
\end{eqnarray}
 This is just the 
 minimal chaotic inflation with quadratic potential. It  predicts 
\begin{eqnarray}
n_S-1\approx-\frac{2}{N}=-0.04\left(\frac{50}{N}\right), ~~~r\approx \frac{8}{N}=0.16\left(\frac{50}{N}\right),  ~~~M\approx \frac{n\!-\!1}{\sqrt{n}}\, 5.1\times  10^{13}\, {\rm GeV},
\end{eqnarray}
which is in  good agreement with the BICEP2 data.  
It is  intriguing that there is no cosmological constant once the fields 
 are settled down to their vacuum although supersymmetry is broken.  
This is due to the  no-scale structure of the K\"ahler potential in Eq. (\ref{K1}) which prevents the appearance of a non-zero vacuum energy, even if supersymmetry is broken \cite{CF}.  
Furthermore, 
the inflaton $\chi\sim(T-\bar{T})$ does not appear in the K\"ahler potential 
which exhibits the global symmetry $T\to T+ i a $ where $a$ is a real constant. This symmetry is not shared by the superpotential (\ref{W1}). However, the theory is natural in the 't Hooft sense as 
 in the $\lambda= 0$ limit, the shift symmetry is recovered. 
The parameter $\lambda$ can be small,  originating from a more fundamental theory where there is a 
small breaking of the shift symmetry.  
As a result, we do not expect higher dimensional operators of the form ${\cal O}_n\sim \chi^n/M_{\rm pl}^{4+n}$ to invalidate the inflationary predictions\cite{yanagida,KLOR}.  However, we should keep in mind that quantum gravity is not expected to respect global symmetries 
so  that quantum gravity effects might generate such operators.

It is interesting to see if there is  a dual theory written in terms of the curvature scalar $R$  and the vector $A_\mu$ as in the $m=0$ Starobinsky case \cite{ADFS}. In terms of $T=\t+i \s$, 
the theory is described by 
\begin{eqnarray}
{\cal L}=\frac{1}{2}R-K_{T\oT}\Big{[}(\partial_\mu \t)^2+(\partial_\mu\s)^2\Big{]}-
\frac{3\lambda}{4}\frac{(\t-f)^2+\s^2}{\t^2\Big{[}1+m (2\t)^{n-1}\Big{]}^2},   \label{l0}
\end{eqnarray}
where 
\begin{eqnarray}
K_{T\oT}=\frac{3}{4}\frac{\Big{[}1+m^2 n (2\t)^{2n-2}-mn(n-3)(2\t)^{n-1}\Big{]}}{\t^2\Big{[}1+m (2\t)^{n-1}\Big{]}^2}.  
\end{eqnarray}
By  performing the conformal transformation 
\begin{eqnarray}
g_{\mu\nu}\to \Omega^2 g_{\mu\nu},
\end{eqnarray}
where \begin{eqnarray}
\Omega^2=f^{-1}\Big{[}y+f+2^{n-1}m\,
(y+f)^n\Big{]}\, , ~~~~~y=\t-f,
\end{eqnarray}
Eq. (\ref{l0}) is written as
\begin{eqnarray}
{\cal L}=\frac{1}{2}\Omega^2\Big{\{}R+6\Omega^{-2}(\partial_\mu \Omega)^2-2 K_{T\oT}
(\partial_\mu y)^2\Big{\}}-\frac{3}{4}\lambda y^2-\frac{3}{4} \lambda \s^2
+A_\mu \partial^\mu \s+\frac{A_\mu A^\mu}{4  K_{T\oT}\Omega^2},
\end{eqnarray}
where we have introduced an auxiliary vector $A_\mu$ \cite{ADFS}. The equations of motion of the latter give back the kinetic term of the $\s$ field. We may also integrate out $\s$ to get 

\begin{eqnarray}
{\cal L}=\frac{1}{2}\Omega^2\Big{\{}R+6\Omega^{-2}(\partial_\mu \Omega)^2-2 K_{T\oT}
(\partial_\mu y)^2\Big{\}}-\frac{3}{4}\lambda y^2+\frac{\nabla_\mu A^\mu}{3 \lambda}
+\frac{A_\mu A^\mu}{4  K_{T\oT}\Omega^2}
\end{eqnarray}
or

\begin{eqnarray}
{\cal L}=\frac{1}{2}\Big{[}y+f+2^{n-1}m\,
(y+f)^n\Big{]}R -\frac{3}{4}\lambda y^2+\frac{\nabla_\mu A^\mu}{3 \lambda}
+\frac{A_\mu A^\mu}{4 K_{T\oT}\Omega^2}-( K_{T\oT}-3 \Omega^{-2}\Omega^2_{,y})
(\partial_\mu y)^2.
\end{eqnarray}
In the zero-momentum limit of the $y$ field ({\it i.e.} during inflation) we may ignore its  derivatives and we can integrate it algebraically. However, although the integration cannot  explicitly be  performed, we can do it perturbative in $m$. The result is 

\begin{align}
{\cal L}&= \frac{1}{2} \left(R+\frac{2}{3}f^2 A_\mu A^\mu\right)+\frac{1}{12 f^2\lambda}
\left(R+\frac{2}{3} f^2A_\mu A^\mu\right)^2+\frac{\nabla_\mu A^\mu}{3 \lambda}\nonumber\\
&+\frac{2^{n-2}}{3^{n+1}(f\lambda)^n}m 
\left(R+\frac{2}{3} f^2 (n^2-3n+1)A_\mu A^\mu\right)
\left(R+\frac{2}{3}f^2A_\mu A^\mu+3f^2 \lambda\right)^n+{\cal O}(m^2),
\end{align}
 where one sees that on the top of the standard $(R+R^2)$ term \cite{ADFS} the leading correction has a maximum higher power of curvature $R^{n+1}$. Thus one can see that coupling the $(R+R^2)$ Starobinsky model leads to the extended Starobinsky model where, among others, an infinite series of the scalar curvature is present.

\section{Conclusions}
In this paper we have reconsidered the prediction of the supersymmetric  Starobinky model of inflation. In its dual formulation, although the real part of the chiral multiplet cannot generate enough tensor modes as an inflaton,  its imaginary part does if appropriate couplings to matter are introduced. 
While its non-supersymmetric version seems to be ruled out by the recent BICEP2 data on the amount of tensor modes,  we have shown  that  the field space of the supersymmetric theory contains  inflationary directions which are in agreement with the current data once appropriate couplings to matter are considered.  The reason is that, along this imaginary direction and once the couplings to matter are considered,  the
model may become  the chaotic single-field model with a quadratic potential.

\section*{Aknowledgements}
We would like to thank M. Porrati    for  enlighting discussions
 as well as R. Kallosh and A. Linde  for pointing out a problem in the v1 of  the paper. 
S.F. is supported by ERC Advanced Investigator Grant n. 226455, ``Supersymmetry, Quantum Gravity and Gauge Fields (Superfelds)".  
The research of A.K. was implemented under the “Aristeia” Action of the “Operational Programme Education and Lifelong Learning” and is co-funded by the European Social Fund (ESF) and National Resources. A.K. is also partially supported by European Union’s Seventh Framework Programme (FP7/2007-2013) under REA grant agreement n. 329083. A.R. is supported by the Swiss National Science Foundation (SNSF), project ``The non-Gaussian Universe" (project number: 200021140236).


\begin{thebibliography}{99}

\bibitem{ade}
  P.~A.~R.~Ade {\it et al.}  [Planck Collaboration],
  arXiv:1303.5082 [astro-ph.CO].

\bibitem{lr}  
D.~H.~Lyth and A.~Riotto,
  Phys.\ Rept.\  {\bf 314}, 1 (1999)
  [hep-ph/9807278].





\bibitem{star} A. A. Starobinsky, Phys. Lett. B {\bf 91}, 99 (1980);
V. F. Mukhanov and G. V. Chibisov, JETP Lett. {\bf 33},
532 (1981) [Pisma Zh. Eksp. Teor. Fiz. {\bf 33}, 549 (1981)];
A. A. Starobinsky, Sov. Astron. Lett. {\bf 9}, 302 (1983).



\bibitem{Whitt:1984pd} 
  B.~Whitt,
  Phys.\ Lett.\ B {\bf 145}, 176 (1984).



%
%
%


  \bibitem{KL} R.~Kallosh and A.~Linde,
  JCAP {\bf 1306}, 028 (2013)
  \bibitem{sugrastaro} 
W.~Buchmuller, V.~Domcke and K.~Kamada,
  Phys.\ Lett.\ B {\bf 726}, 467 (2013)
  [arXiv:1306.3471 [hep-th]]
  \bibitem{ENO}
  J.~Ellis, D.~V.~Nanopoulos and K.~A.~Olive,
  Phys.\ Rev.\ Lett.\  {\bf 111}, 111301 (2013)
  [arXiv:1305.1247 [hep-th]]; 
  J.~Ellis, D.~V.~Nanopoulos and K.~A.~Olive,
  JCAP {\bf 1310}, 009 (2013)
  [arXiv:1307.3537]; 
  \bibitem{FKLP}
   S.~Ferrara, R.~Kallosh, A.~Linde and M.~Porrati,
  Phys. Rev. D 88 (2013)085038, 
  arXiv:1307.7696 [hep-th];
   S.~Ferrara, R.~Kallosh, A.~Linde and M.~Porrati,
  JCAP {\bf 1311}, 046 (2013)
  [arXiv:1309.1085 [hep-th]].
  
  \bibitem{FKR}
F.~Farakos, A.~Kehagias and A.~Riotto,
  Nucl.\ Phys.\ B {\bf 876}, 187 (2013)
  [arXiv:1307.1137];
  \bibitem{FKD} A.~Kehagias, A.~M.~Dizgah and A.~Riotto,
  Phys.\ Rev.\ D {\bf 89}, 043527 (2014)
  
\bibitem{KT} 
S. V. Ketov and A. A. Starobinsky,  Phys.
Rev. D {\bf 83}, 063512 (2011) [arXiv:1011.0240 [hep-th]]; 
S. V. Ketov and A. A. Starobinsky,
 JCAP {\bf 1208}, 022 (2012) [arXiv:1203.0805 [hep-th]]; 
S. V. Ketov and S. Tsujikawa, Phys. Rev. D {\bf 86}, 023529 (2012)
[arXiv:1205.2918 [hep-th]];
  S.~V.~Ketov and T.~Terada,
  JHEP {\bf 1312}, 040 (2013)
  [arXiv:1309.7494 [hep-th]].
  
\bibitem{fv}
 S.Ferrara, M.T.Grisaru and P.Van Nieuwenhuizen, Nucl. Phys. B 138(1978) 430.








\bibitem{Cecotti:1987sa} 
  S.~Cecotti,
  Phys.\ Lett.\ B {\bf 190}, 86 (1987).

\bibitem{CFPS} 
  S.~Cecotti, S.~Ferrara, M.~Porrati and S.~Sabharwal,
  Nucl.\ Phys.\ B {\bf 306}, 160 (1988).

\bibitem{bicep}
  P.~A.~R.~Ade {\it et al.}  [BICEP2 Collaboration],
  arXiv:1403.3985 [astro-ph.CO].

\bibitem{chaotic} A. Linde, Phys. Lett B {\bf 129}, 177(1983).

\bibitem{Kallosh:2014ona}  
  R.~Kallosh,
  arXiv:1402.0527 [hep-th]; 
  R.~Kallosh,
  arXiv:1402.0527 [hep-th];
 A.~Linde,
  arXiv:1402.0526 [hep-th].

\bibitem{KLr} 
  R.~Kallosh, A.~Linde, B.~Vercnocke and W.~Chemissany,
  arXiv:1403.7189 [hep-th].

\bibitem{jap} K.~Hamaguchi, T.~Moroi and T.~Terada,
  arXiv:1403.7521 [hep-ph].

\bibitem{CK} 
  S.~Cecotti and R.~Kallosh,
  arXiv:1403.2932 [hep-th].  
  
\bibitem{CFG}
  S.~Cecotti, S.~Ferrara and L.~Girardello,
  Nucl.\ Phys.\ B {\bf 294} (1987) 537.
  
\bibitem{FFS} 
  S.~Ferrara, P.~Fre and A.~S.~Sorin,
  arXiv:1401.1201 [hep-th]; arXiv:1311.5059 [hep-th].
  
\bibitem{KLR} 
  R.~Kallosh, A.~Linde and D.~Roest,
  JHEP {\bf 1311}, 198 (2013)
  [arXiv:1311.0472 [hep-th]].
  
  \bibitem{F0} 
  D.~Z.~Freedman,
  Phys.\ Rev.\ D {\bf 15}, 1173 (1977).

\bibitem{ADFS} 
  I.~Antoniadis, E.~Dudas, S.~Ferrara and A.~Sagnotti,
  arXiv:1403.3269 [hep-th].

\bibitem{goldstino}
  M.~Rocek,
  Phys.\ Rev.\ Lett.\  {\bf 41} (1978) 451;
  U.~Lindstrom, M.~Rocek,
  Phys.\ Rev.\  D {\bf 19} (1979) 2300;
  R.~Casalbuoni, S.~De Curtis, D.~Dominici, F.~Feruglio and R.~Gatto,
  Phys.\ Lett.\  B {\bf 220} (1989) 569;
Z.~Komargodski and N.~Seiberg,
JHEP {\bf 0909} (2009) 066 [arXiv:0907.2441 [hep-th]].

\bibitem{luis} 
  L.~Alvarez-Gaume, C.~Gomez and R.~Jimenez,
  Phys.\ Lett.\ B {\bf 690}, 68 (2010)
  [arXiv:1001.0010 [hep-th]]; JCAP {\bf 1103}, 027 (2011)
  [arXiv:1101.4948 [hep-th]].

\bibitem{SUGRA} For a review see:
D.~Z.~Freedman and A.~Van Proeyen,
  Cambridge, UK: Cambridge Univ. Pr. (2012) 607 p.


  
\bibitem{fkvp}
S.~Ferrara, R.~Kallosh and A.~Van Proeyen,
  JHEP {\bf 1311} (2013) 134
  [arXiv:1309.4052 [hep-th]].





  




\bibitem{FKP}  S.~Ferrara, A.~Kehagias and M.~Porrati, 
  Phys.\ Lett.\ B {\bf 727}, 314 (2013)
  [arXiv:1310.0399 [hep-th]]; 


\bibitem{CF}
  E.~Cremmer, S.~Ferrara, C.~Kounnas and D.~V.~Nanopoulos,
  Phys.\ Lett.\ B {\bf 133} (1983) 61; 
   J.~R.~Ellis, A.~B.~Lahanas, D.~V.~Nanopoulos and K.~Tamvakis,
  Phys.\ Lett.\ B {\bf 134} (1984) 429.  arXiv:1403.3985 [astro-ph.CO].

  \bibitem{yanagida} 
  M.~Kawasaki, M.~Yamaguchi and T.~Yanagida,
  Phys.\ Rev.\ Lett.\  {\bf 85}, 3572 (2000)
  [hep-ph/0004243].

\bibitem{KLOR} 
  R.~Kallosh, A.~Linde, K.~A.~Olive and T.~Rube,
  Phys.\ Rev.\ D {\bf 84}, 083519 (2011)
  [arXiv:1106.6025 [hep-th]].




\bibitem{EKN} 
  J.~R.~Ellis, C.~Kounnas and D.~V.~Nanopoulos,
  Nucl.\ Phys.\ B {\bf 241}, 406 (1984).

 \end{thebibliography}
\end{document}